\documentclass[11pt,a4paper]{article}
\pdfoutput=1

\usepackage{jheppub_kim}
\usepackage{rotating}
\usepackage{graphicx,epsfig}
\usepackage{amsmath}
\usepackage {amssymb}
\usepackage{subfigure}

\usepackage{relsize}

\def\beq{\begin{equation}}
\def\eeq{\end{equation}}
\def\bea{\begin{eqnarray}}
\def\eea{\end{eqnarray}}

\begin{document}

\title{New observational constraints on $f(R)$ gravity from cosmic chronometers}

\author[a]{Rafael C. Nunes}

\author[b]{Supriya Pan}

\author[c,d]{Emmanuel N. Saridakis}

\author[e,a]{Everton M. C. Abreu}

\affiliation[a]{Departamento de F\'isica, Universidade Federal de Juiz de Fora,
36036-330,
Juiz de
Fora, MG, Brazil}

\affiliation[b]{Department of Physical Sciences, Indian Institute of Science Education
and Research -- Kolkata, Mohanpur -- 741246, West Bengal, India}

\affiliation[c]{Physics Division,
National Technical University of Athens, 15780 Zografou Campus,
Athens, Greece}

\affiliation[d]{CASPER, Physics Department, Baylor University, Waco, TX 76798-7310, USA}

\affiliation[e]{Grupo de F\' isica Te\'orica e Matem\'atica F\' isica, Universidade 
Federal 
Rural do 
Rio de Janeiro, 23890-971, Serop\'edica, RJ, Brazil}

\emailAdd{rcnunes@fisica.ufjf.br}
\emailAdd{span@iiserkol.ac.in}
\emailAdd{Emmanuel$_-$Saridakis@baylor.edu}
\emailAdd{evertonabreu@ufrrj.br}

\abstract{
We use the recently released cosmic chronometer data and the  latest measured value of 
the local Hubble parameter, combined with  the latest joint light curves of  Supernovae 
Type Ia,  and Baryon Acoustic Oscillation distance
measurements, in order to impose constraints on the viable and most used $f(R)$ gravity
models. We consider four $f(R)$ models, namely the Hu-Sawicki, the Starobinsky, the 
Tsujikawa, and the exponential one, and we parametrize them introducing a distortion 
parameter $b$ that quantifies the deviation from $\Lambda$CDM cosmology. Our analysis 
reveals that a small but non-zero deviation from $\Lambda$CDM cosmology is slightly 
favored, with the corresponding fittings exhibiting very efficient $AIC$ and $BIC$ 
Information Criteria values. Clearly, $f(R)$ gravity is consistent with observations, and 
it can serve as a candidate for modified gravity.}


\maketitle

\section{Introduction}

According to the concordance  model of cosmology the universe must have experienced 
two accelerated expansion phases, at early and late times respectively. This behavior 
cannot be reproduced within the standard framework of general relativity and Standard 
Model of particles, and thus extra degrees of freedom should be introduced. Firstly, one 
can attribute these extra degrees of freedom to new, exotic forms of matter, such as the 
inflaton field at early times (for reviews see \cite{Olive:1989nu,Bartolo:2004if}) and/or 
the dark energy concept at late times (for reviews see \cite{Copeland:2006wr,Cai:2009zp}).
Alternatively, one can consider the extra degrees of freedom to have gravitational 
origin, i.e. to arise from a gravitational modification that possesses general relativity 
as a particular limit (see \cite{Nojiri:2006ri,Capozziello:2011et} and references 
therein). Note that the latter approach has the additional advantage that it might 
improve renormalizability and thus alleviate the difficulties towards   
quantization 
\cite{Stelle:1976gc,Biswas:2011ar}.

In the usual approach to gravitational modification one adds higher-order corrections to  
the Einstein-Hilbert action. The simplest such modification arises from extending the 
Ricci 
scalar $R$ to an arbitrary function $f(R)$, which can lead to interesting behavior 
at early times \cite{Starobinsky:1980te}, as well as explain the late-time acceleration
\cite{DeFelice:2010aj,Capozziello:2002rd,Capozziello:2003tk, Carroll:2003wy, 
Nojiri:2003ni, 
Capozziello:2005ku, Das:2005bn, Nojiri:2006gh, Miranda:2009rs, Campista:2010jb, 
Mukherjee:2014fna, 
Paliathanasis:2016tch}, or describe both phases in a unified 
way \cite{Nojiri:2010wj,Nojiri:2003ft, Nojiri:2007cq}, and can explain the large scale 
structure 
distribution 
in the universe \cite{Santos:2016vdv, Voivodic:2016kog}. However, one can 
construct many other classes of curvature-based modified gravities, such as  $f(G)$ 
gravity \cite{Nojiri:2005jg, DeFelice:2008wz}, Lovelock 
gravity \cite{Lovelock:1971yv, Deruelle:1989fj}, Weyl gravity
\cite{Mannheim:1988dj, Flanagan:2006ra}, Galileon theory 
\cite{Nicolis:2008in, Deffayet:2009wt, Deffayet:2009mn, Leon:2012mt}, or even extend to 
torsion-based modifications, such as  $f(T)$ gravity  
\cite{Cai:2015emx, Ben09, Linder:2010py, Chen:2010va, Paliathanasis:2016vsw, 
Nunes:2016qyp},   $f(T,T_G)$ gravity
\cite{Kofinas:2014owa, Kofinas:2014aka, Kofinas:2014daa}, etc.

An important and probably the most justified question in all gravitational modifications 
is what is the choice of the involved arbitrary function. A first constraining of the 
possible forms comes from theoretical arguments, such that the requirement for a 
ghost-free theory that possesses stable perturbations \cite{DeFelice:2010aj}, or the 
desire for the theory to possess Noether symmetries \cite{Paliathanasis:2011jq, 
Paliathanasis:2015aos}. However, 
in order to further 
constrain the remaining huge class of theories the main tool is the use of 
observational data and the requirement for a successful reproduction of the universe 
history, as well as of the local/solar system behavior. In the case of $f(R)$ gravity 
such a confrontation with cosmological data was performed in  
  \cite{Hwang:2001pu} (using data from cosmic microwave background (CMB) probes), 
in \cite{Yamamoto:2010ie,Abebe:2013zua} (using  Large Scale Structure (LSS) data),
  in \cite{Arapoglu:2010rz} (using neutron stars mass-radius data),  
in \cite{Aviles:2012ir}  (using Supernovae type Ia  (SNe Ia) and  Hubble Parameter data), 
in \cite{Capozziello:2003gx} (using SNe Ia and  CMB data),
 in \cite{Fay:2007gg,Santos:2008qp} (using SNe Ia, CMB and 
data from baryonic acoustic oscillations (BAO) probes), in \cite{Dev:2008rx} (using 
Hubble Parameter and BAO data),
in \cite{Carvalho:2008am} (using  CMB, BAO, Hubble Parameter data),
in \cite{Amarzguioui:2005zq,Song:2007da,Basilakos:2013nfa,Basilakos:2016nyg}
  (using SNe Ia, CMB, and  growth rate data),
  in \cite{Schmidt:2009am}  (using SNe Ia, CMB, BAO, Hubble Parameter and cluster 
abundance constraints), and  
  in \cite{Lombriser:2010mp} (using SNe Ia, CMB, BAO, Hubble Parameter, gravitational 
lenses and  growth rate data). Additionally, the comparison with solar system data 
was performed in \cite{Ruggiero:2006qv,Chiba:2006jp,Amendola:2007nt,Nojiri:2007as, 
Capozziello:2007eu,Iorio:2016sqy,Hu:2007nk}. Hence, constructions that pass all the above 
constraints are called viable models.

In the present work we intend to provide updated observational constraints on $f(R)$ 
gravity models, using the latest released cosmic chronometer data set 
and the latest released local value of the Hubble parameter with 2.4\% 
precision, along with the standard cosmological 
probes for dark energy analysis, such as Supernovae type Ia and baryonic acoustic 
oscillations data. In particular, we will consider four viable $f(R)$ 
models, namely (i) Hu-Sawicki model, (ii) Starobinsky model, (iii) Tsujikawa model and 
finally (iv) exponential $f(R)$ model, and we will provide the updated constraints 
and contour plots for the involved parameters. The plan of the manuscript is the 
following: In section \ref{fRcosmology} we briefly review 
$f(R)$ gravity and its cosmological application, focusing on four viable $f(R)$ 
models. In section \ref{data} we describe the data sets used 
for the observational confrontation, while in section \ref{results} we provide the 
results 
of our analysis, namely the updated observational 
constraints on the various model parameters and observational quantities. Finally, we 
close our work in section 
\ref{Conclusions}, with a summary and discussion.

\section{$f(R)$ gravity and cosmology}
\label{fRcosmology}

In this section we briefly review $f(R)$ gravity and we proceed to its cosmological 
application. Then we examine four specific $f(R)$ models, which, amongst the variety of 
$f(R)$ scenarios, pass the basic theoretical and observational tests and thus they are 
considered as  viable ones.

\subsection{$f(R)$ gravity}

In $f(R)$ gravitational theories one extends the Einstein-Hilbert action to 
\begin{equation}
S = \int d^4 x \sqrt{-g}\,\, \frac{f(R)}{16\pi G} +S_m+S_r\,,
\label{action0}
\end{equation}
with $R$  the Ricci scalar and $G$ the gravitational constant, and where we have also
considered the actions for the matter and radiation sectors, $S_m$ and $S_r$
respectively. Following the metric formulation, variation of the action (\ref{action0})
with respect to the metric $g_{\mu\nu}$ leads to the field equations
\begin{eqnarray}
&&\!\!\!\!\!\!\!\!\!\!\!
F G_{\mu\nu}
=
-\frac{1}{2} g_{\mu \nu} \left( FR - f \right)
+ \nabla_{\mu}\nabla_{\nu}F -g_{\mu \nu} \Box F   +
8\pi G\,  \left[T^{(m)}_{\mu \nu} +T^{(r)}_{\mu \nu}\right]
\,,
\label{eoms0}
\end{eqnarray}
with 
$G_{\mu\nu}=R_{\mu\nu}-\left(1/2\right)g_{\mu\nu}R$ 
the Einstein tensor, ${\nabla}_{\mu}$   the covariant derivative,
$\Box \equiv g^{\mu \nu} {\nabla}_{\mu} {\nabla}_{\nu}$, and where we have defined
$F(R) \equiv f_{,R}= d f(R)/dR$. Additionally, $ T^{(m)}_{\mu \nu} $ and $ T^{(r)}_{\mu 
\nu} $ are respectively  the  energy-momentum tensors for the matter and radiation 
sectors,
corresponding to $S_m$ and $S_r$.

Before proceeding, and for completeness, we mention that apart from the above 
metric (or second order) formulation of $f(R)$ gravity, in which the field equations are 
derived through variation of the action with respect to the metric tensor, and where the 
affine connection depends only on the metric, one could have the Palatini  (or first 
order) formulation, where the metric and the connection are treated as independent 
variables in the action variation, under the assumption that the matter part of the action 
does not depend on the connection \cite{DeFelice:2010aj}. For a general $f(R)$ form these 
two approaches lead to different field equations, and only in the General Relativity case, 
i.e for $f(R)=R$, the two formulations coincide. Finally, one could also have the 
metric-affine formulation, in which the Palatini variation is used but without the 
additional assumption that the matter action is connection-independent (the metric-affine 
formulation reduces to metric or Palatini formulations if extra considerations are made). 
In the present work we focus on the standard metric formulation, since Palatini formalism 
might exhibit difficulties in being compatible with observations and experiments, as well 
as it faces problems with the formulation of the Cauchy problem due to the presence of  
matter fields  higher-derivatives  in the field equations (see 
\cite{DeFelice:2010aj} and the references therein).

\subsection{$f(R)$ cosmology}

We now proceed to the cosmological application of $f(R)$ gravity. Hence, we consider the
usual homogeneous and isotropic geometry, characterized by the Friedmann-Robertson-Walker
(FRW) background metric
\begin{equation}
ds^2=-dt^2+a(t)^2\left[\frac{dr^2}{(1-kr^2)}+r^2(d\theta^2+\sin^2\theta
\,d\phi^2)\right],
\end{equation}
with $a(t)$ the scale factor and $k$ the
spatial curvature (with $k=0,-1,+1$ for flat, open and closed universe
respectively). Focusing for simplicity to the flat case, and inserting the FRW 
metric into the field equations (\ref{eoms0}), we obtain the modified
Friedmann equations
\begin{eqnarray}
3FH^2
=
8\pi G  \left(\rho_m+\rho_r\right) +\frac{1}{2} \left( FR - f \right)
-3H\dot{F}\,,
\label{FR1a} \\
-2F\dot{H}
=
8\pi G  \left( \rho_m + P_m +\rho_r + P_r \right)
+\ddot{F}-H\dot{F}\,,
\label{FR2a}
\end{eqnarray}
where
$H\equiv\dot{a}/a$ is the Hubble parameter, with dot denoting derivatives with respect to
the cosmic time $t$. Furthermore, we have considered that the matter and radiation sectors
correspond to perfect fluids with energy densities $\rho_m$, $\rho_r$ and pressures
$P_m$, $P_r$ respectively. Finally, note that in flat FRW geometry one obtains the useful
relation
\begin{eqnarray}
\label{Ricciscalar}
R=6\left(2H^2+\dot{H}\right).
\end{eqnarray}

Observing the form of the Friedmann equations (\ref{FR1a}), (\ref{FR2a}),  and comparing 
to
the usual ones, namely $3H^2=8\pi G  \left(\rho_m +\rho_r+\rho_{DE}\right)$ as well as
$-2\dot{H}=8\pi G  \left( \rho_m + P_m + \rho_r+ P_r+ \rho_{DE} + P_{DE}\right) $,
we deduce that in the scenario at hand we obtain an effective dark energy
sector, with dark energy density and pressure defined as 
\begin{eqnarray}
&&\rho_{DE}\equiv
\frac{1}{8\pi G } \left[
\frac{1}{2} \left( FR - f \right)
-3H \dot{F}
+3\left(1-F\right)H^2
\right]
, \ \ \ \ \label{rhoDDE}\\
\label{pDE}
&&P_{DE}\equiv \frac{1}{8\pi G }
\Big[
-\frac{1}{2} \left( FR - f \right)
+\ddot{F}+2H \dot{F}  
-\left(1-F\right)\left(2\dot{H}+3H^2\right)
\Big]
,
\end{eqnarray}
while its effective equation-of-state parameter reads:
\begin{eqnarray}
\label{wfT}
w \equiv P_{DE}/\rho_{DE}.
\end{eqnarray}
One can easily see that $\rho_{DE}$ and $P_{DE}$ defined in (\ref{rhoDDE}), (\ref{pDE}) 
satisfy the usual evolution equation
\begin{eqnarray}
\dot{\rho}_{DE}+3H(\rho_{DE}+P_{DE})=0.
\end{eqnarray}
Finally, the equations close considering the standard matter and radiation evolution 
equations, namely
\begin{eqnarray}
&&\dot{\rho}_{m}+3H(\rho_{m}+P_{m})=0,\\
&&\dot{\rho}_{r}+3H(\rho_{r}+P_{r})=0,
\end{eqnarray}
respectively.

\subsection{Specific $f(R)$ models }
\label{manymodels}

In this subsection we review the most used and viable $f(R)$ models. First of all, a 
given 
$f(R)$ model must satisfy some basic theoretical constraints, namely to possess a 
positive  effective gravitational constant, as well as to exhibit stable cosmological 
perturbations \cite{DeFelice:2010aj}. In particular, one should have 
\begin{eqnarray}
f_{,R}>0\ \ \text{for}\ \ 
R\geq R_0,
\end{eqnarray}
  with $R_0$ the present value of the Ricci scalar, in order to avoid a ghost 
state, and 
\begin{eqnarray}
f_{,RR}>0\ \ \text{for}\ \ 
R\geq R_0,
\end{eqnarray}
 in order to avoid the scalar-field degree of 
freedom to become tachyonic. Additionally, a given $f(R)$ model must satisfy some basic 
observational requirements. Specifically, one should have
\begin{eqnarray}
f(R)\rightarrow R-2\Lambda\ \ \text{for}\ \ 
R\geq R_0,
\end{eqnarray}
  in order to be able to reproduce the matter era and to obtain 
consistency with equivalence principle and local gravity constraints, and
\begin{eqnarray}
0<\frac{R f_{,RR}}{f_{,R}}(r)<1 \ \ \text{at}\ \ 
r=-\frac{Rf_{,R}}{f}=-2,
\end{eqnarray}
 in order to have the presence and stability of a late-time   de 
Sitter solution \cite{DeFelice:2010aj}. Hence, considering viable models that have up to 
two parameters, one can write them as 
\begin{eqnarray}
f(R)=R-2\Lambda y(R,b),
\label{fRyRb1}
\end{eqnarray}
where the function $y(R,b)$ quantifies the deviation from Einstein gravity, i.e. the 
effect of the $f(R)$ modification, through the distortion
parameter $b$.

Having these in mind, one can construct four viable $f(R)$ models, that have been 
investigated in detail in the literature, which are given below.

\begin{enumerate}

\item The Hu-Sawicki model \cite{Hu:2007nk}.

This model corresponds to 
\begin{eqnarray}
f(R)=
R -
\frac{c_1 R_{\mathrm{HS}} \left(R/R_{\mathrm{HS}}\right)^p}{c_2
\left(R/R_{\mathrm{HS}}\right)^p + 1},
\label{HSfR}
\end{eqnarray}
where $c_1$, $c_2$ and $R_{\mathrm{HS}}$ are   parameters  and $p>0$ a positive constant. 
Note that not all of these parameters are independent since, using the first Friedmann 
equation 
(\ref{FR1a}) at present, one of them can be eliminated in favor of the present values of 
the density parameters  $\Omega_{i0}=\frac{8\pi 
G \rho_{i0}}{3H_0^2}$ as well as the present value of the Hubble function $H_0$  (the 
subscript ``0'' denotes the current value of a quantity). One can easily rewrite 
(\ref{HSfR}) to the form (\ref{fRyRb1}), with
\begin{equation}
y(R, b) = 1- \frac{1}{1+ \Bigl(\frac{R}{\Lambda b} \Bigr)^p},
\end{equation}
where the two free model parameters read as 
$\Lambda = \frac{c_1R_{\mathrm{HS}}}{2c_2} $ and $b= 2 
c_2^{1-1/p}/c_1$. Hence, one can see that for $b\rightarrow0$ (i.e. for 
$c_1\rightarrow\infty$, $R_{\mathrm{HS}}\rightarrow 0$, with 
$c_1 R_{\mathrm{HS}}\rightarrow 2c_2\Lambda$) the Hu-Sawicki model reduces to 
$\Lambda$CDM cosmology since $f(R)\rightarrow R-2\Lambda$.
 We mention here that the above reduction/mapping to two parameters (plus $p$) 
offers an effective way in order to be able to investigate the fittings on all parameters 
through a reconstruction method via error propagation. In principle one could try to fit 
all parameters independently, however the existing data (in terms of quantity and 
precision) cannot lead to a good precision fits.

\item
The
 Starobinsky model \cite{Starobinsky:2007hu}.

This model corresponds to 
\begin{eqnarray}
f(R)=
R -
\lambda R_{\mathrm{S}} \left[ 1 -
\left(1+\frac{R^2}{R_{\mathrm{S}}^2} \right)^{-n}
\right],
\label{StarfR}
\end{eqnarray}
with $\lambda (>0)$ and $R_{\mathrm{S}}$ the free parameters and $n>0$ a positive 
constant. One can rewrite 
(\ref{StarfR}) to the form (\ref{fRyRb1}), with
\begin{equation}
y(R, b) = 1- \frac{1}{\Bigl[1+ \left(\frac{R}{\Lambda \, b} \right)^2\Bigr]^n},
\end{equation}
where $\Lambda=  \lambda R_{\mathrm{S}}/2$ and $b= 2/\lambda$. Thus,  for $b\rightarrow0$ 
(i.e. for 
$\lambda\rightarrow\infty$, $R_{\mathrm{S}}\rightarrow 0$, with 
$\lambda R_{\mathrm{S}}\rightarrow 2\Lambda$) the Starobinsky model reduces to 
$\Lambda$CDM cosmology, namely $f(R)\rightarrow R-2\Lambda$. Note that the mapping 
to two parameters (plus $n$) is performed similarly to the previous Hu-Sawicki model.

\item
The Tsujikawa model \cite{Tsujikawa:2007xu}.

This model corresponds to 
\begin{eqnarray}
f(R)=
R - \mu R_{\mathrm{T}}
\tanh\left( \frac{R}{R_{\mathrm{T}}} \right),
\label{TsusfR}
\end{eqnarray}
where $\mu (>0)$ and $R_{\mathrm{T}} (>0)$ are two positive constants.
 One can rewrite 
(\ref{TsusfR}) as (\ref{fRyRb1}), defining
\begin{equation}
y(R, b) = \tanh \left( \frac{R}{b \Lambda} \right),
\end{equation}
where $\Lambda= \mu R_{\mathrm{T}}/ 2$, and $b= 2/\mu$. The Tsujikawa model reduces to 
$\Lambda$CDM cosmology for $b\rightarrow0$ 
(i.e. for 
$\mu\rightarrow\infty$, $R_{\mathrm{T}}\rightarrow 0$, with 
$\mu R_{\mathrm{T}}\rightarrow 2\Lambda$).

\item
The exponential gravity model  \cite{Cognola:2007zu,Elizalde:2010ts,Zhang:2005vt}.

This case corresponds to 
\begin{eqnarray}
f(R)=
R -\beta R_{\mathrm{E}}\left(1-e^{-R/R_{\mathrm{E}}}\right),
\label{ExponfR}
\end{eqnarray} 
with $\beta$, $R_{\mathrm{E}}$ the model parameters.
One can rewrite 
(\ref{ExponfR}) to the form (\ref{fRyRb1}), with
\begin{equation}
y(R, b) = 1- e^{-R/(\Lambda\, b)},
\end{equation}
where $\Lambda = \beta R_{\mathrm{E}}/ 2$, and $b= 2/\beta$.  This  model reduces to 
$\Lambda$CDM cosmology for   $b\rightarrow0$ 
(i.e. for 
$\beta\rightarrow\infty$, $R_{\mathrm{E}}\rightarrow 0$, with 
$\beta R_{\mathrm{E}}\rightarrow 2\Lambda$).

\end{enumerate}

\section{Current Observational Data}
\label{data}

In  this work  we are interested in constraining $f(R)$ gravity using observational data 
acquired from probes that map the expansion history of the late-time universe, namely 
lying in the redshift region $z < 2.36$. The main ingredient of our analysis is the 
Hubble 
parameter measurements obtained with the cosmic chronometers (CC) technique, which are 
the 
latest and   model-independent measurements of the Hubble parameter, and thus
provide better constraints on a cosmological model. In addition, we consider standard
probes such as Supernovae Type Ia (SNe Ia), local Hubble parameter value $H_0$ ones, and
Baryon Acoustic Oscillation  (BAO) distance measurements, in order to reduce the
degeneracy between the free parameters of the models. We mention here that 
it would be both interesting and necessary to try to constrain $f(R)$ gravity on 
smaller scales, too. Although at galaxies and smaller scales the effect of modified 
gravity is expected to be very small and hardly detectable, indeed at galaxy clusters it 
might lead to observational constraints. This interesting subject lies beyond the scope 
of the present work, and it is left for a future project.  The following subsections 
describe 
the employed data sets for our analysis.

\subsection{Cosmic chronometer dataset and local value of the Hubble constant}
\label{cc-data}

The Cosmic Chronometer (CC) approach is a very powerful implementation in understanding 
the 
universe evolution. It was first introduced in \cite{cc1}, and the method determines
the Hubble parameter data through the differential age evolution of the passively 
evolving early-type galaxies. Since the Hubble parameter for FRW universe can be 
expressed as $H= -(1+z)^{-1} dz/dt$, by measuring the quantity $dz/dt$, one can directly 
measure the Hubble parameter data. Hence, the CC data are very powerful in order to 
provide better constraints on cosmological models. For a detailed description on the 
implementation of CC data, all possible kind of uncertainties, as well as some related 
issues, we refer the reader to \cite{cc2}. Here we consider the compilation of Hubble 
parameter measurements as provided in \cite{cc3, cc2}. The data set contains 30 $H(z)$ 
measurements  \cite{cc2,cc3,cc4,cc5,cc6} obtained through the CC approach in the redshift 
range $0 < z <2$, and it roughly covers about 10 Gyr of cosmic time. Moreover, in 
addition 
to the CC data, in our investigation we include the new local value of $H_0$ as measured 
by \cite{riess} with a 2.4 $\%$ determination, which yields $H_0 = 73.02 \pm 1.79$ 
km s${}^{-1}$ 
Mpc${}^{-1}$.

\subsection{Type Ia Supernovae}
\label{snia-data}

SNe Ia provided the first signal for a universe acceleration  \cite{snia1,snia2}, and
they serve as the main observational data set to probe the late-time, dark-energy 
epoch. In this work we consider the latest ``joint light curves'' (JLA) sample 
\cite{snia3} containing 740 SNe Ia in the redshift range $z \in [0.01, 1.30]$. From the 
observational point of view, the  distance modulus of a SNe Ia can be abstracted from its 
light curve, assuming that supernovae with identical color, shape and galactic 
environment, have on average the same intrinsic luminosity for all redshifts. This 
hypothesis is quantified by an empirical linear relation, yielding a standardized 
distance modulus $\mu = 5
\log_{10}(d_L/10pc)$ of the form
\begin{align}
\label{snia1}
 \mu = m^\ast_B - (M_B - \alpha \times X_1 + \beta \times C),
\end{align}
where $m^\ast_B$ corresponds to the observed peak magnitude in rest frame B band and
$\alpha$, $\beta$, and $M_B$ are nuisance parameters in the distance estimate.
The absolute magnitude is related to the host stellar mass ($M_{stellar}$) by a simple
step function:
$M_B = M_B$ if $M_{stellar} < 10^{10} M_\odot $, otherwise $M_B = M_B + \Delta_M$. The
light-curve parameters ($m^\ast_B$, $X_1$ and $C$) result from the fit of a model of
the SNe Ia spectral sequence to the photometric data. In our analysis we assume $M_B$,
$\Delta_M$, $\alpha$ and $\beta$ as nuisance parameters.

\subsection{Baryon Acoustic oscillation}
\label{bao-data}

Another potential cosmological test comes from the baryon acoustic oscillations (BAO) 
data. In our analysis, we  adopt the  following  BAO  data  to  constrain  the  expansion
history of the universe: the  measurement from  the  Six Degree  Field  Galaxy
Survey (6dF) \cite{bao1}, the  Main Galaxy Sample  of  Data  Release 7  of  Sloan
Digital  Sky  Survey  (SDSS-MGS) \cite{bao2}, the  LOWZ  and  CMASS  galaxy  samples  of
the Baryon  Oscillation  Spectroscopic  Survey  (BOSS-LOWZ  and  BOSS-CMASS,
respectively) \cite{bao3},  and the distribution of the LymanForest in BOSS (BOSS-Ly)
\cite{bao4}. These measurements and their corresponding effective redshift $z$ are
summarized in Table \ref{tab1}.
\begin{table}[!h]
      \begin{center}
          \begin{tabular}{ccccc}
          \hline
          \hline
           Survey &  $z$     &  Parameter   &  Measurement  & Reference  \\
          \hline
 6dF             & 0.106 & $r_s/D_V$  &  0.327 $\pm$ 0.015 & \cite{bao1} \\
 SDSS-MGS        & 0.10  &  $D_V/r_s$ &  4.47 $\pm$ 0.16   & \cite{bao2} \\
 BOSS-LOWZ       & 0.32  & $D_V/r_s$  &  8.47  $\pm$ 0.17  & \cite{bao3} \\
 BOSS-CMASS      & 0.57  &  $D_V/r_s$ &  13.77  $\pm$ 0.13 & \cite{bao3} \\
 BOSS-$Ly_{\alpha}$& 2.36& $c/(H r_s)$&  9.0    $\pm$ 0.3  & \cite{bao4} \\
 BOSS-$Ly_{\alpha}$& 2.36 & $D_A/r_s$ &  10.08  $\pm$ 0.4  & \cite{bao4} \\
          \hline
          \hline
          \end{tabular}
      \end{center}
      \caption{\textit{Baryon acoustic oscillation (BAO) data measurements included in our
analysis.}}
      \label{tab1}
\end{table}

\section{Observational Constraints}
\label{results}

In this section we shall present the main observational constraints, extracted for the 
four viable $f(R)$ models reviewed in subsection \ref{manymodels}. We use the data 
described in the previous section, and we first perform fittings using  Cosmic 
Chronometer (CC)  $+$ $H_0$ observations. Then, we proceed to the combination of all data 
sets, namely of  SNe Ia ``joint light curves'' (JLA) $+$ BAO $+$ CC $+$ $H_0$. 
To fit the free parameters in these $f(R)$ scenarios we use the publicly available code 
CLASS \cite{class} in the interface with the public Monte Carlo code Monte Python 
\cite{monte}. Moreover, in our analysis we use the Metropolis Hastings algorithm as our 
sampling method. In the following subsections we shall separately discuss the 
observational results on the various $f(R)$ models.

\subsection{Constraints on Hu-Sawicki model}
\label{results-HS}

\begin{figure*}[ht]
\includegraphics[width=5.5in, height=5.3in]{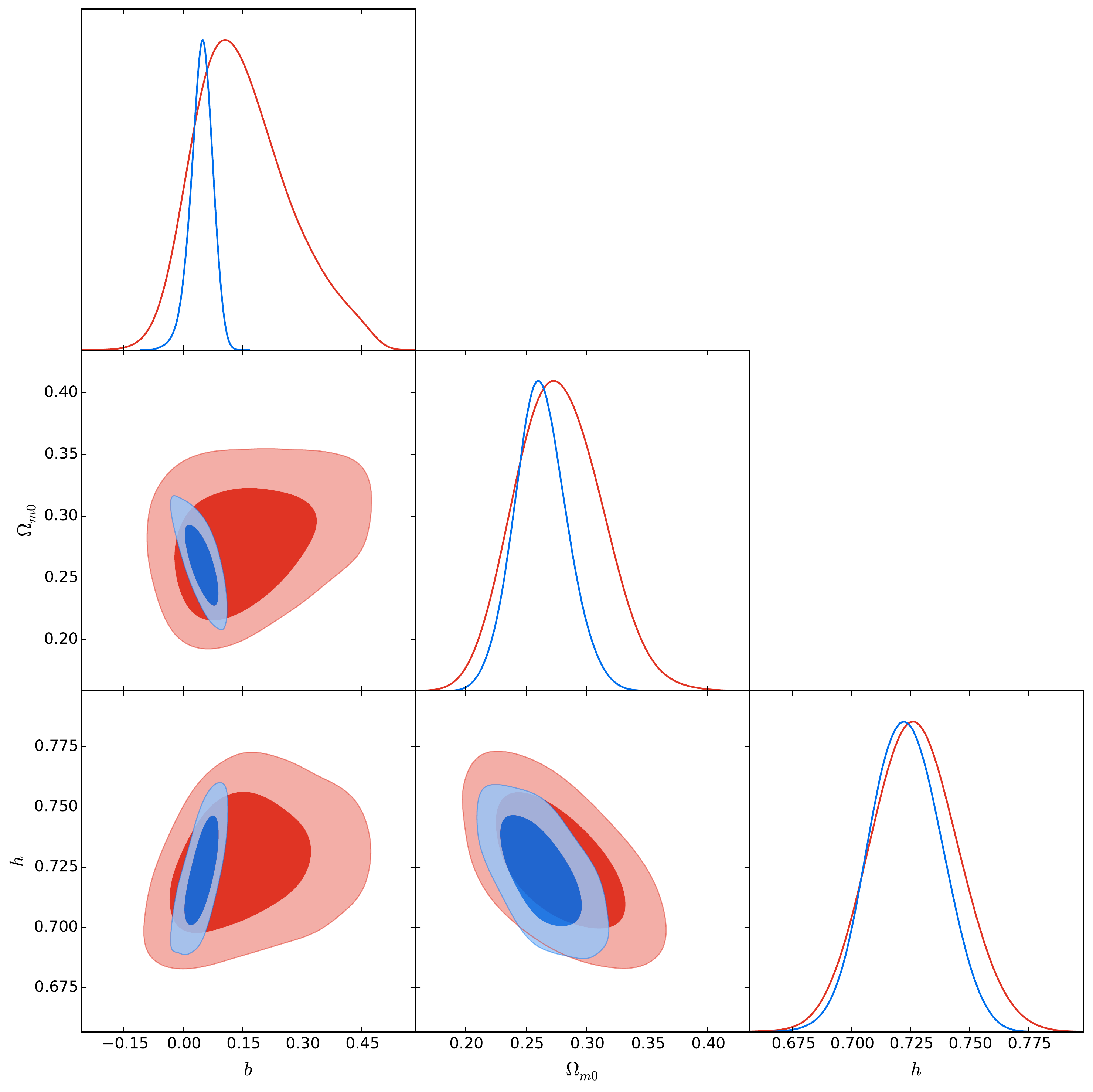}
\caption{\label{fig1-HS} \textit{
Contour plots for the free parameter $b$, as well as for the present value of the matter 
density parameter $\Omega_{m0}$ and for the dimensionless Hubble parameter $h$, for the  
Hu-Sawicki model of (\ref{HSfR}). The red and pink regions correspond to 1$\sigma$ and 
2$\sigma$ confidence level in the case of  CC $+H_0$ data sets, while the blue and light 
blue regions   correspond to 1$\sigma$ and 
2$\sigma$ confidence level for the combined analysis of JLA $+$ BAO $+$ CC 
$+$ $H_0$ data sets. Additionally, we present the corresponding marginalized
one-dimensional posterior distributions. The parameter $\Omega_m$ includes both baryons 
and cold dark matter, i.e. $\Omega_m= \Omega_{cdm}+ \Omega_b$,
and $h= H_0/100$  km s${}^{-1}$ 
Mpc${}^{-1}$. }}
\end{figure*}

We fit the Hu-Sawicki model of (\ref{HSfR}), following the above procedure, 
and in Fig. \ref{fig1-HS} we present the contour plots of various quantities and model 
parameters, for both used data sets.  We mention that since there is a known 
degeneracy between $p$ and $\Omega_{m0}$ that requires to fix $p$  a priori, we choose the 
case $p=1$ since it is  the most used case in the literature \cite{Basilakos:2013nfa} (in 
principle one could perform the fittings for higher  $p$ too, nevertheless
higher values have difficulties in fitting the data).
Additionally, in 
Table \ref{tab-HS}  we summarize 
the 
best fit values of the data analysis for the two data sets respectively. As we observe, 
and interestingly enough, the parameter $b$ which quantifies the deviation from 
$\Lambda$CDM cosmology is favored to have nonzero values for both data sets (for the 
combined analysis,  i.e. for JLA $+$ BAO $+$ CC $+$ $H_0$, the contours come closer to 
zero comparing to the CC $+H_0$ case, but the zero value is only marginally allowed), 
although the zero value is still inside the allowed region at both 1$\sigma$ and 
2$\sigma$ 
confidence level. Hence, observations seem to slightly favor a small but not  
non-zero deviation from  $\Lambda$CDM cosmology. This is one of the main results of the
present work. Although some indications towards this direction were previously obtained  
in \cite{Basilakos:2013nfa}, in the present work, with the addition of CC data, this 
behavior is enhanced.

\begin{table}[ht]
\begin{center}
\begin{tabular}{|r|c|c|c|c|c|c|c|c|c|c|}

\hline \small
 \footnotesize Parameters& \multicolumn{1}{c|}{ \footnotesize CC$+$ 
$H_0$}&\multicolumn{1}{c|}{\footnotesize JLA $+$ BAO $+$ CC $+$ $H_0$}\\
\cline{2-3}
\cline{4-5}
\cline{5-6}
\cline{6-7}
\cline{7-8}
\cline{8-9}
&\footnotesize Best fit $\pm$ $1\sigma$ $\pm$ $2\sigma$  &\footnotesize Best fit $\pm$ 
$1\sigma$ $\pm$ $2\sigma$\\
\hline
\footnotesize $b$&{\footnotesize 
$0.107^{+0.316+0.393}_{-0.158-0.221}$}&{\footnotesize $0.
048^{+0.062+0.081}_{-0.077-0.125}$}\\
\footnotesize $h$ &{\footnotesize 
$0.729^{+0.034+0.047}_{-0.034-0.049}$}&{\footnotesize $0.
722^{+0.042+0.044}_{-0.033-0.047}$}\\
\footnotesize $\Omega_m$ &{\footnotesize 
$0.264^{+0.069+0.102}_{-0.058-0.083}$}&{\footnotesize 
$0.264^{+0.059+0.078}_{-0.055-0.066}$}\\
\hline
\end{tabular}
\caption{\label{tab-HS}Summary of the best fit values and main results for the free 
parameter $b$, as well as 
for the present value of the matter density parameter $\Omega_{m0}$ and for the 
dimensionless Hubble parameter $h$, for the  Hu-Sawicki model of (\ref{HSfR}), using   
CC$+$ $H_0$ and 
JLA$+$BAO$+$CC $+H_0$ observational data. The parameter
$\Omega_m$ includes
both baryons and cold dark matter,
i.e. $\Omega_m= \Omega_{cdm}+ \Omega_b$,
and $h= H_0/100$  km s${}^{-1}$ 
Mpc${}^{-1}$.}
\end{center}
\end{table}

\subsection{Constraints on Starobinsky model}
\label{results-STA}

\begin{figure*}[ht]
   \includegraphics[width=5.5in, height=5.3in]{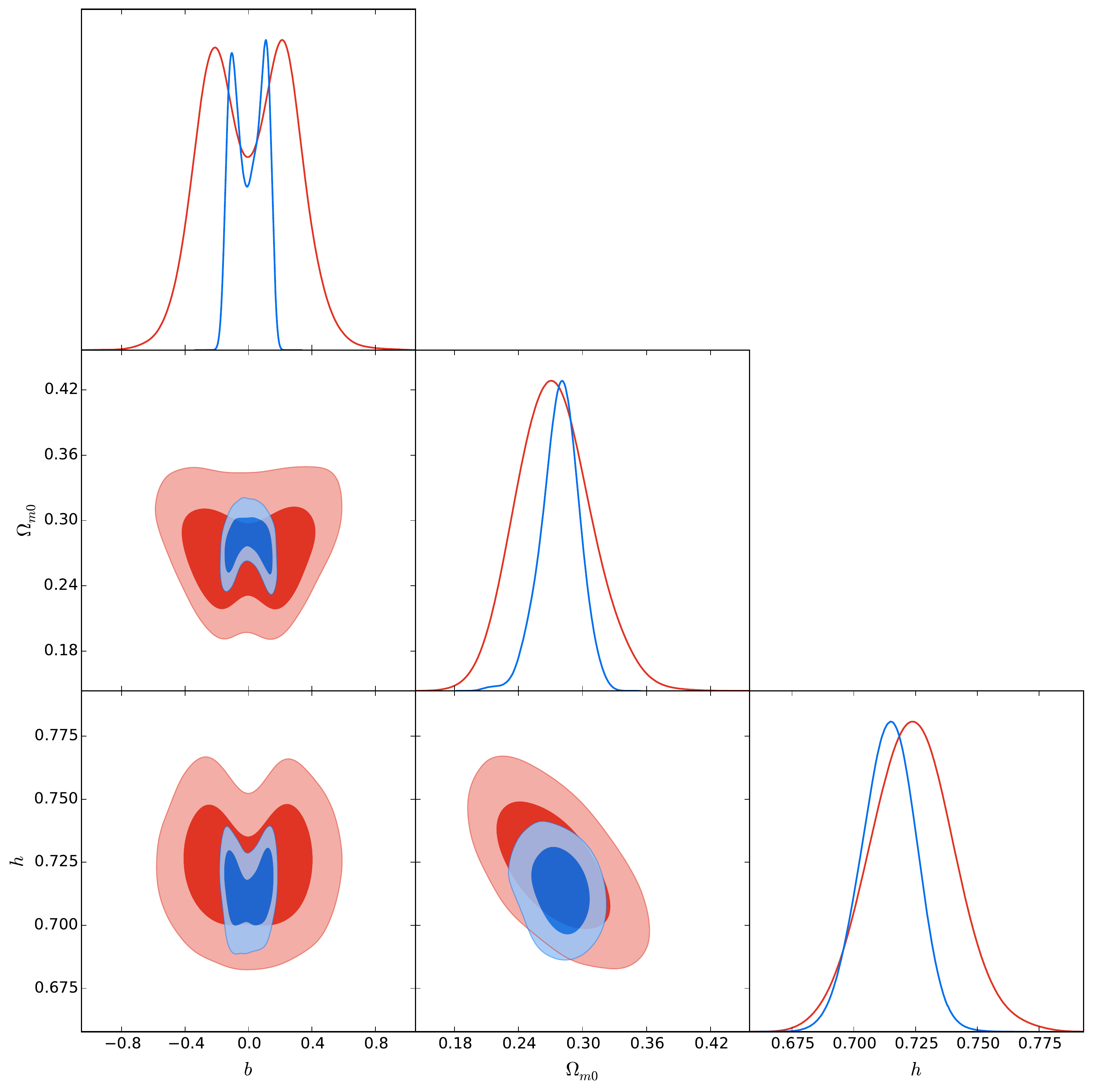}
   \caption{\label{fig1-STA} 
   \textit{Contour plots for the free parameter $b$, as well as for the present value of 
the matter 
density parameter $\Omega_{m0}$ and for the dimensionless Hubble parameter $h$, for the  
Starobinsky model of (\ref{StarfR}). The red and pink regions correspond to 1$\sigma$ and 
2$\sigma$ confidence level in the case of  CC $+H_0$ data sets, while the blue and light 
blue regions   correspond to 1$\sigma$ and 
2$\sigma$ confidence level for the combined analysis of JLA $+$ BAO $+$ CC 
$+$ $H_0$ data sets. Additionally, we present the corresponding marginalized
one-dimensional posterior distributions. The parameter $\Omega_m$ includes both baryons 
and cold dark matter, i.e. $\Omega_m= \Omega_{cdm}+ \Omega_b$,
and $h= H_0/100$  km s${}^{-1}$ 
Mpc${}^{-1}$.}}
\end{figure*}

For the case of Starobinsky model of (\ref{StarfR}) with $n= 1$ (similarly to the 
  Hu-Sawicki model there is a degeneracy between $n$ and $\Omega_{m0}$ that 
requires to fix $n$  a priori, and we choose the 
value $n=1$ since it is  the most used case in the literature), and similarly to 
the 
previous 
model, 
we perform the fittings using two different data sets, namely CC $+$ $H_0$ data, and  JLA 
$+$ BAO $+$ CC $+$ $H_0$. In Fig. \ref{fig1-STA} we depict the contour plots of various 
quantities, while in Table \ref{tab-STA} we provide the corresponding best fit values. 
\begin{table}[ht]
\begin{center}
\begin{tabular}{|r|c|c|c|c|c|c|c|c|c|c|}
\hline \small
 \footnotesize Parameters& \multicolumn{1}{c|}{ \footnotesize CC$+$ 
$H_0$}&\multicolumn{1}{c|}{\footnotesize JLA $+$ BAO $+$ CC $+$ $H_0$}\\
\cline{2-3}
\cline{4-5}
\cline{5-6}
\cline{6-7}
\cline{7-8}
\cline{8-9}
&\footnotesize Best fit $\pm$ $1\sigma$ $\pm$ $2\sigma$  &\footnotesize Best fit $\pm$ 
$1\sigma$ $\pm$ $2\sigma$\\
\hline
\footnotesize $b$ &{\footnotesize 
$0.229^{+0.254+0.438}_{-0.710-0.893}$}&{\footnotesize $0.
111^{+0.070+0.074}_{-0.286-0.304}$}\\
\footnotesize $h$ &{\footnotesize 
$0.727^{+0.031+0.047}_{-0.031-0.047}$}&{\footnotesize $0.
714^{+0.030+0.038}_{-0.028-0.033}$}\\
\footnotesize $\Omega_m$ &{\footnotesize 
$0.261^{+0.065+0.102}_{-0.055-0.080}$}&{\footnotesize 
$0.269^{+0.050+0.061}_{-0.042-0.054}$}\\
\hline
\end{tabular}
\caption{\label{tab-STA}
Summary of the best fit values and main results for the free parameter $b$, as well as 
for the present value of the matter density parameter $\Omega_{m0}$ and for the 
dimensionless Hubble parameter $h$, for the Starobinsky model of (\ref{StarfR}), using   
CC$+$ $H_0$ and 
JLA$+$BAO$+
$CC$+$ $H_0$ observational data. The parameter
$\Omega_m$ includes
both baryons and cold dark matter,
i.e. $\Omega_m= \Omega_{cdm}+ \Omega_b$,
and $h= H_0/100$  km s${}^{-1}$ 
Mpc${}^{-1}$.}
\end{center}
\end{table}
In 
this model, the distortion parameter $b$ which quantifies the deviation of the model from 
$\Lambda$CDM cosmology has a slight preference to be non zero (as can be especially seen 
by the marginalized one-dimensional posterior distribution), however the zero value is 
clearly allowed, at both 1$\sigma$ and 2$\sigma$ confidence level, and hence
this model can observationally coincide with $\Lambda$CDM scenario.

\subsection{Constraints on Tsujikawa model}
\label{results-TSU}

\begin{figure*}
   \includegraphics[width=5.5in, height=5.3in]{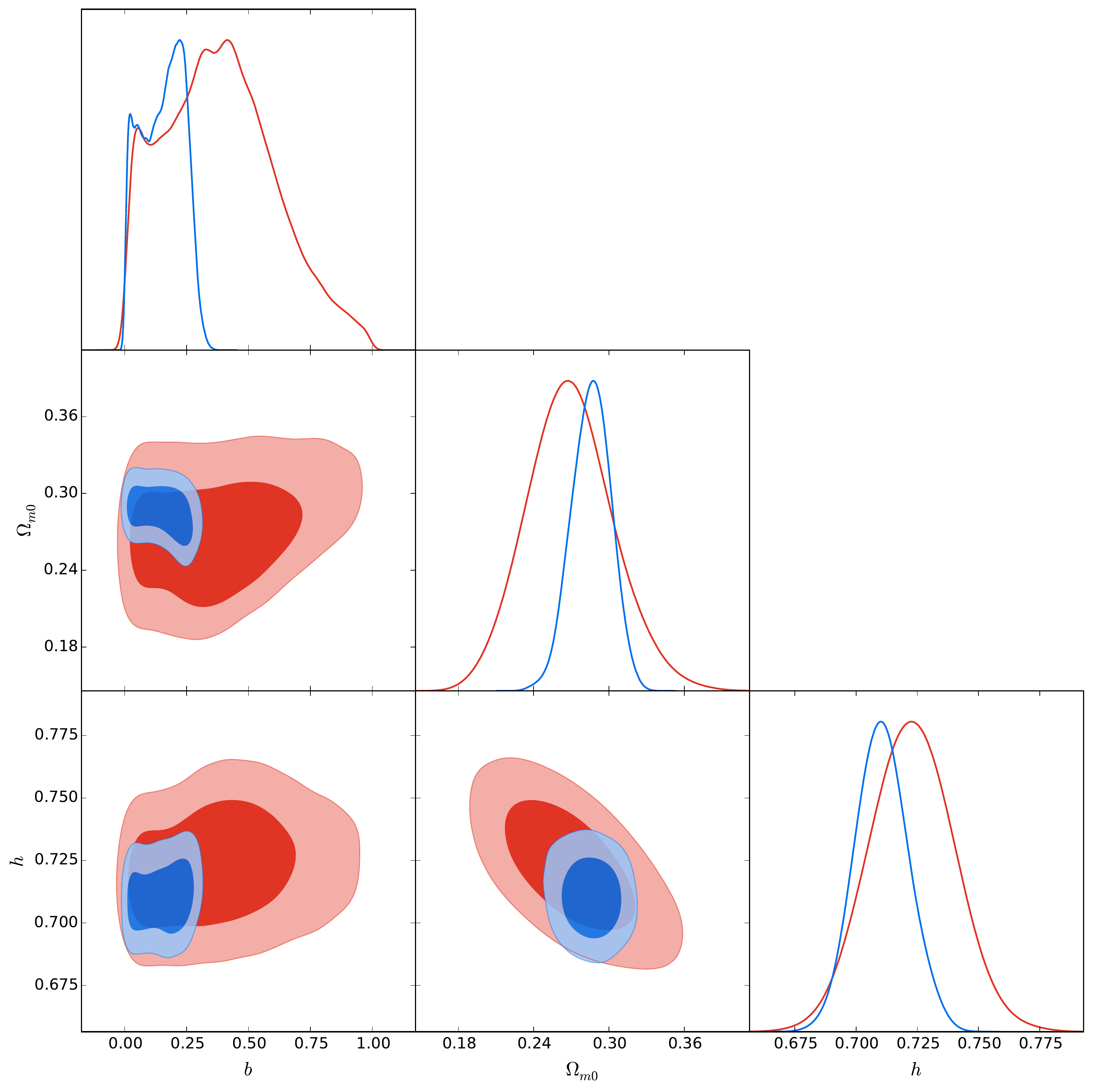}
   \caption{\label{fig1-TSU} \textit{Contour plots for the free parameter $b$, as well as 
for the present value of 
the matter 
density parameter $\Omega_{m0}$ and for the dimensionless Hubble parameter $h$, for the  
Tsujikawa model of (\ref{TsusfR}). The red and pink regions correspond to 1$\sigma$ and 
2$\sigma$ confidence level in the case of  CC $+H_0$ data sets, while the blue and light 
blue regions   correspond to 1$\sigma$ and 
2$\sigma$ confidence level for the combined analysis of JLA $+$ BAO $+$ CC 
$+$ $H_0$ data sets. Additionally, we present the corresponding marginalized
one-dimensional posterior distributions. The parameter $\Omega_m$ includes both baryons 
and cold dark matter, i.e. $\Omega_m= \Omega_{cdm}+ \Omega_b$,
and $h= H_0/100$  km s${}^{-1}$ 
Mpc${}^{-1}$.
 }}
\end{figure*}
\begin{figure*}[ht]
   \includegraphics[width=5.5in, height=5.3in]{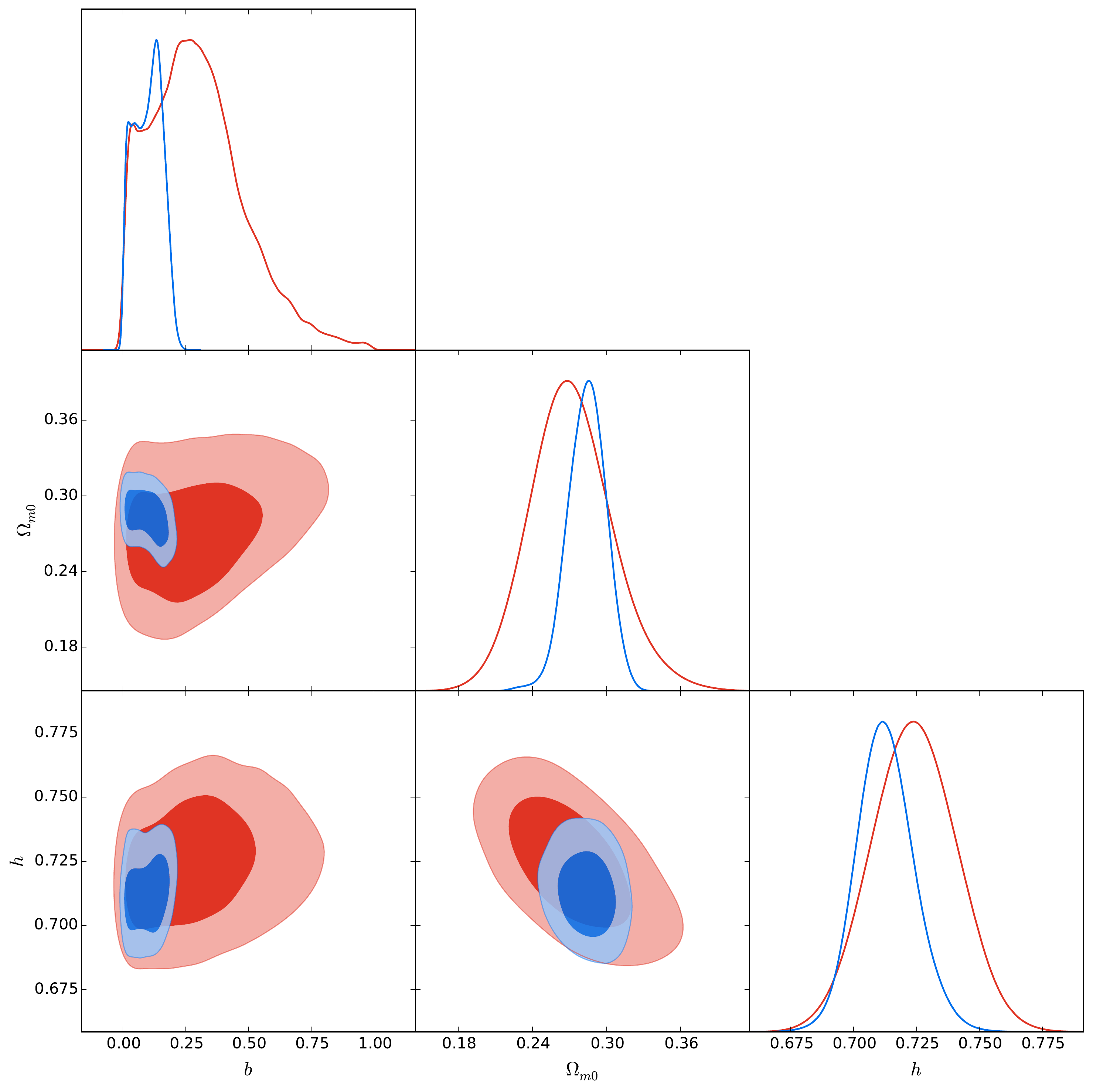}
   \caption{\label{fig1-EXP} \textit{
   Contour plots for the free parameter $b$, as well as 
for the present value of 
the matter 
density parameter $\Omega_{m0}$ and for the dimensionless Hubble parameter $h$, for the  
exponential $f(R)$ gravity model of (\ref{ExponfR}. The red and pink regions correspond 
to 
1$\sigma$ and 
2$\sigma$ confidence level in the case of  CC $+H_0$ data sets, while the blue and light 
blue regions   correspond to 1$\sigma$ and 
2$\sigma$ confidence level for the combined analysis of JLA $+$ BAO $+$ CC 
$+$ $H_0$ data sets. Additionally, we present the corresponding marginalized
one-dimensional posterior distributions. The parameter $\Omega_m$ includes both baryons 
and cold dark matter, i.e. $\Omega_m= \Omega_{cdm}+ \Omega_b$,
and $h= H_0/100$  km s${}^{-1}$ 
Mpc${}^{-1}$. }}
\end{figure*}  
For the case of Tsujikawa model of (\ref{TsusfR}), in Fig. \ref{fig1-TSU} we depict the 
contour plots  arisen from the fitting analysis, while in Table  \ref{tab-TSU} we present 
the corresponding best fit values for both used data sets, namely for    
CC$+$ $H_0$ and for
JLA$+$BAO$+$CC $+H_0$ observational data.
\begin{table}[ht]
\begin{center}
\begin{tabular}{|r|c|c|c|c|c|c|c|c|c|c|}
\hline \small
 \footnotesize Parameters& \multicolumn{1}{c|}{ \footnotesize CC$+$ 
$H_0$}&\multicolumn{1}{c|}{\footnotesize JLA $+$ BAO $+$ CC $+$ $H_0$}\\
\cline{2-3}
\cline{4-5}
\cline{5-6}
\cline{6-7}
\cline{7-8}
\cline{8-9}
&\footnotesize Best fit $\pm$ $1\sigma$ $\pm$ $2\sigma$  &\footnotesize Best fit $\pm$ 
$1\sigma$ $\pm$ $2\sigma$\\
\hline
\footnotesize $b$ &{\footnotesize 
$0.425^{+0.400+0.575}_{-0.424-0.424}$}&{\footnotesize $0.
196^{+0.124+0.154}_{-0.195-0.195}$}\\
\footnotesize $h$ &{\footnotesize 
$0.726^{+0.031+0.047}_{-0.031-0.047}$}&{\footnotesize $0.
709^{+0.031+0.035}_{-0.025-0.035}$}\\
\footnotesize $\Omega_m$ &{\footnotesize 
$0.261^{+0.063+0.098}_{-0.056-0.081}$}&{\footnotesize 
$0.284^{+0.041+0.048}_{-0.044-0.052}$}\\
\hline
\end{tabular}
\caption{\label{tab-TSU}
Summary of the best fit values and main results for the free 
parameter $b$, as well as 
for the present value of the matter density parameter $\Omega_{m0}$ and for the 
dimensionless Hubble parameter $h$, for the Tsujikawa model of (\ref{TsusfR}), using   
CC$+$ $H_0$ and 
JLA$+$BAO$+
$CC$+$ $H_0$ observational data. The parameter
$\Omega_m$ includes
both baryons and cold dark matter,
i.e. $\Omega_m= \Omega_{cdm}+ \Omega_b$,
and $h= H_0/100$  km s${}^{-1}$ 
Mpc${}^{-1}$.   }
\end{center}
\end{table}
As we observe, in this case 
the distortion parameter $b$ is clearly non-zero, with the zero value only very 
marginally allowed. Thus, Tsujikawa model exhibits an observable deviation from 
$\Lambda$CDM cosmology. This is one of the main results of the present work.

\subsection{Constraints on exponential model}
\label{results-EXP}

For the case of exponential $f(R)$ gravity model of (\ref{ExponfR}), in Fig. 
\ref{fig1-EXP} we present the likelihood contours arisen from the fitting analysis, while 
in Table \ref{tab-EXP} we provide the corresponding best fit values for both used data 
sets, namely for CC $+$ $H_0$ and JLA $+$ BAO $+$ CC $+$ $H_0$. 
Similarly to the previous 
model, the parameter $b$ that quantifies the deviation from $\Lambda$CDM cosmology is 
clearly non-zero, with the zero value only very marginally allowed.
Hence, exponential 
$f(R)$ gravity could be observationally distinguished from $\Lambda$CDM paradigm. 
Furthermore, note that this scenario exhibits a very similar behavior with Tsujikawa 
model, which was expected due to the relation of the hyperbolic tangent with the 
exponentials.
\begin{table}[ht]
\begin{center}
\begin{tabular}{|r|c|c|c|c|c|c|c|c|c|c|}
\hline \small
 \footnotesize Parameters& \multicolumn{1}{c|}{ \footnotesize CC$+$ 
$H_0$}&\multicolumn{1}{c|}{\footnotesize JLA $+$ BAO $+$ CC $+$ $H_0$}\\
\cline{2-3}
\cline{4-5}
\cline{5-6}
\cline{6-7}
\cline{7-8}
\cline{8-9}
&\footnotesize Best fit $\pm$ $1\sigma$ $\pm$ $2\sigma$ & \footnotesize Best fit $\pm$ 
$1\sigma$ $\pm$ $2\sigma$\\
\hline
\footnotesize $b$ &{\footnotesize 
$0.289^{+0.341+0.635}_{-0.289-0.289}$}&{\footnotesize $0.
130^{+0.089+0.118}_{-0.130-0.130}$}\\
\footnotesize $h$  &{\footnotesize 
$0.727^{+0.031+0.046}_{-0.032-0.047}$}&{\footnotesize $0.
711^{+0.030+0.039}_{-0.026-0.033}$}\\
\footnotesize $\Omega_m$ &{\footnotesize $0.261^{+0.064+0.100}_{-0.055-0.080}$} 
&{\footnotesize
 $0.284^{+0.040+0.043}_{-0.049-0.062}$}\\
\hline
\end{tabular}
\caption{\label{tab-EXP}
Summary of the best fit values and main results for the free 
parameter $b$, as well as 
for the present value of the matter density parameter $\Omega_{m0}$ and for the 
dimensionless Hubble parameter $h$, for the exponential $f(R)$ gravity model of 
(\ref{ExponfR}), using   
CC$+$ $H_0$ and 
JLA$+$BAO$+
$CC$+$ $H_0$ observational data. The parameter
$\Omega_m$ includes
both baryons and cold dark matter,
i.e. $\Omega_m= \Omega_{cdm}+ \Omega_b$,
and $h= H_0/100$  km s${}^{-1}$ 
Mpc${}^{-1}$.}
\end{center}
\end{table}

\subsection{Model comparison}
\label{sec-model-selection}

We close the observational analysis with the present subsection, in which we compare 
the fittings of the various models, using the standard information criteria. There are 
two 
main such criteria, namely the 
 Akaike Information Criterion ($AIC$) \cite{Akaike1974} and the
Bayesian or Schwarz Information Criterion
($BIC$) \cite{bic}. These are respectively defined as
\begin{eqnarray}
&& AIC  = -2 \ln \mathcal{L}+ 2 d = \chi^2_{min} + 2 d,\label{aic}
\end{eqnarray}
and
\begin{eqnarray}
&& BIC  = -2 \ln \mathcal{L}+  d \ln N = \chi^2_{min} +  d \ln N,\label{bic}
\end{eqnarray}
where $\mathcal{L}= \exp\left(-\chi_{min}^2/2\right)$ is the maximum likelihood function, 
$d$ is the number of model parameters and $N$ denotes the total number of data points 
used 
in the statistical analysis. Definitely, one must also introduce a reference scenario, 
with respect of which the comparisons will be performed, and obviously this is
$\Lambda$CDM cosmology. Hence, for any given model denoted by $M$, and calculating     
the difference $\Delta X= X_{M}- X_{\Lambda CDM}$ (where $X= AIC$ or $BIC$),
one may result to the following conclusions \cite{dunsby}: (i) If $\Delta X \leq 2$, 
then the concerned model has substantial support with respect to the reference model 
(i.e. it has evidence to be a good cosmological model),
(ii) if $4 \leq \Delta X \leq 7$ it is an indication for less support with respect to 
the reference model, and finally, (iii) if $\Delta X \geq 10$ then the model 
has no observational support. Note that including the nuisance parameters 
arising from Supernoave Type Ia, we have $6$ model parameters in $\Lambda$CDM paradigm,  
while in all $f(R)$ models we have $7$ free parameters.

In Table \ref{tab-aic-bic} we present the values of $\Delta X$ for the four analyzed 
models, for both used data sets, namely for CC $+$ $H_0$ and JLA $+$ BAO $+$ CC $+$ 
$H_0$ ones. As we can see, for both data sets $\Delta AIC \leq 2$, and hence    
these models are very efficient and in  very good agreement with observations, and they 
fit the data slightly better than $\Lambda$CDM paradigm. Concerning  $\Delta BIC$, we 
observe that it acquires slightly larger values, and therefore according to this 
criterion $\Lambda$CDM scenario is slightly favored, although all $f(R)$ models are still 
very efficient. In summary, we deduce that all models behave very efficiently, and 
especially the Hu-Sawicki and Starobinsky ones seem to have a better fitting behavior 
comparing to $\Lambda$CDM paradigm.

\begin{table*}[ht]
\begin{center}
\begin{tabular}{|r|c|c|c|c|c|c|c|c|c|c|}

\hline \small
 \footnotesize Models & \multicolumn{4}{c|}{ \footnotesize CC$+$ 
$H_0$}&\multicolumn{4}{c|}{\footnotesize JLA $+$ BAO $+$ CC $+$ $H_0$}\\
\cline{2-3}
\cline{4-5}
\cline{5-6}
\cline{6-7}
\cline{7-8}
\cline{8-9}
&\footnotesize $AIC$ &\footnotesize $\Delta AIC$ &\footnotesize $BIC$ & \footnotesize 
$\Delta BIC$ &
\footnotesize $AIC$ &\footnotesize $\Delta AIC$ & \footnotesize $BIC$ & \footnotesize 
$\Delta BIC$\\
\hline
\footnotesize $\Lambda$CDM Model&{\footnotesize $28.205$}&{\footnotesize 
$0$}&{\footnotesize $36.
809$}&{\footnotesize $0$}&{\footnotesize $721.084$}&{\footnotesize $0$}&{\footnotesize 
$749.017$}&{\footnotesize $0$}\\
\footnotesize Hu-Sawicki Model &{\footnotesize $28.744$}&{\footnotesize 
$0.539$}&{\footnotesize $38.782$}&{\footnotesize $1.973$}&{\footnotesize 
$720.840$}&{\footnotesize 
$-0.244$}&{\footnotesize $753.428$}&{\footnotesize $4.411$}\\
\footnotesize Starobinsky Model &{\footnotesize $29.096$}&{\footnotesize 
$0.891$}&{\footnotesize 
$39.134$}&{\footnotesize $2.325$}&{\footnotesize $721.726$}&{\footnotesize 
$0.642$}&{\footnotesize 
$754.314$}&{\footnotesize $5.297$}\\
\footnotesize Tsujikawa Model &{\footnotesize $29.407$}&{\footnotesize 
$1.202$}&{\footnotesize $39.445$}&{\footnotesize $2.636$}&{\footnotesize 
$722.966$}&{\footnotesize 
$1.882$}&{\footnotesize $755.554$}&{\footnotesize $6.537$}\\
\footnotesize Exponential Model &{\footnotesize $29.310$}&{\footnotesize 
$1.105$}&{\footnotesize 
$39.347$}&{\footnotesize $2.538$}&{\footnotesize $722.548$}&{\footnotesize 
$1.464$}&{\footnotesize 
$755.136$}&{\footnotesize $6.119$}\\
\hline
\end{tabular}
\caption{\label{tab-aic-bic}
Summary of the $AIC$ and $BIC$ values, as well as of their difference from the reference 
model of $\Lambda$CDM cosmology, for the CC$+$ $H_0$ and 
JLA$+$BAO$+$CC$+$ $H_0$ data sets, for all four analyzed $f(R)$ models.}
\end{center}
\end{table*}

\section{Conclusions}
\label{Conclusions}

In this manuscript we have implemented  the recently released cosmic chronometer 
data in order to impose constraints on the viable and most used $f(R)$ gravity models. In 
particular, we used  the recent cosmic chronometer data set,
along with the latest measured value of the local Hubble parameter, $H_0 = 73.02 \pm
1.79$ km s${}^{-1}$ Mpc${}^{-1}$ \cite{riess}, while we additionally performed a combined 
analysis using  the latest ``joint light curves'' (JLA) SNe Ia sample 
\cite{snia3}  in the redshift range $z \in [0.01, 1.30]$, as well as 
baryon acoustic oscillation (BAO) data points from various probes.

We examined four specific $f(R)$ models, namely the Hu-Sawicki, the Starobinsky, the 
Tsujikawa, and the exponential one, and we parametrized them introducing a distortion 
parameter $b$ that quantifies the deviation from $\Lambda$CDM cosmology. Thus, we used 
the above observational data in order to fit this parameter,  along with various other
cosmological quantities.

For the Hu-Sawicki scenario the parameter $b$ is favored to have nonzero values for both 
data sets, although the zero value is still inside the allowed region at both 1$\sigma$ 
and 2$\sigma$ confidence level, and thus a small but not non-zero deviation from  
$\Lambda$CDM cosmology is slightly favored. For the Starobinsky scenario $b$ has a slight 
preference to be non zero, however the zero value is clearly allowed, at both 1$\sigma$ 
and 2$\sigma$ confidence levels, and hence this model can observationally coincide with 
$\Lambda$CDM scenario. However, for the Tsujikawa and exponential models the distortion 
parameter $b$ is clearly non-zero, with the zero value only very  marginally allowed. 
Hence, both these models exhibit an observable deviation from $\Lambda$CDM cosmology. 
This 
is one of the main results of the present work. Note that although some indications 
towards this direction had been previously obtained in the literature, in the present 
work, with the addition of CC data, this behavior is much more clear.

Finally, we performed a comparison of the fitting procedure with $\Lambda$CDM 
paradigm, using the $AIC$ and $BIC$ Information Criteria. According to $AIC$, for both 
data sets all four $f(R)$ models are very efficient and sightly better than $\Lambda$CDM  
one, while according to  $BIC$ the $\Lambda$CDM scenario is slightly better, 
nevertheless with all $f(R)$ models quite efficient.
 
In summary, using for the first time the recently released cosmic chronometer data, 
combined with data from other probes, we fitted the viable and most used  $f(R)$ gravity 
models. As we saw, clearly $f(R)$ gravity is consistent with observations.  
Additionally, a small but non-zero deviation from $\Lambda$CDM cosmology is slightly 
favored, with the corresponding fittings exhibiting very efficient information criteria 
values. These features indicate that $f(R)$ gravity may serve as a good candidate for 
gravitational modifications.  \\

\section*{Acknowledgments}
S.P. 
acknowledges 
Science and Engineering Research Board (SERB), Govt. of India, for awarding
National Post-Doctoral Fellowship (File No: PDF/2015/000640). E.M.C.A. thanks CNPq 
(Conselho Nacional de Desenvolvimento Cient\'ifico e Tecnol\'ogico), and the Brazilian 
scientific support federal agency, for partial financial support, under 
Grants numbers 302155/2015-5, 302156/2015-1 and 442369/2014-0 and the hospitality of 
Theoretical Physics Department at Federal University of Rio de Janeiro (UFRJ), where part 
of this work was carried out. This article is 
based upon work from COST Action ``Cosmology and Astrophysics Network
for Theoretical Advances and Training Actions'', supported by COST (European Cooperation
in Science and Technology).

\end{document}